\documentclass[aps,twocolumn,pra,showpacs,superscriptaddress]{revtex4}

\usepackage{graphicx}
\usepackage{amsfonts}
\usepackage{amssymb,amsmath}
\usepackage{verbatim}

\newcommand{\ket}[1]{\left| #1 \right>}

\newcommand{\void}[1]{}

\begin{document}

\title{Entanglement creation in a quantum dot-nanocavity system}

\author{Ralf Blattmann}
\affiliation{Institut f{\"u}r Physik, Universit{\"a}t Augsburg,
Universit{\"a}tsstra{\ss}e 1, D-86153 Augsburg, Germany}
\affiliation{Nanosystems Initiative Munich, Schellingstr. 4, D-80799 M\"{u}nchen, Germany}

\author{Hubert J. Krenner}
\affiliation{Institut f{\"u}r Physik, Universit{\"a}t Augsburg,
Universit{\"a}tsstra{\ss}e 1, D-86153 Augsburg, Germany}
\affiliation{Nanosystems Initiative Munich, Schellingstr. 4, D-80799 M\"{u}nchen, Germany}
\affiliation{Center for NanoScience ({\it CeNS}), Ludwig-Maximilians-Universit{\"a}t M{\"u}̈nchen, Geschwister-Scholl-Platz 1, D-80539 M\"{u}nchen, Germany}

\author{Sigmund Kohler}
\affiliation{Instituto de Ciencia Materiales de Madrid, CSIC, Cantoblanco, E-28049 Madrid, Spain}

\author{Peter H{\"a}nggi}
\affiliation{Institut f{\"u}r Physik, Universit{\"a}t Augsburg,
Universit{\"a}tsstra{\ss}e 1, D-86153 Augsburg, Germany}
\affiliation{Nanosystems Initiative Munich, Schellingstr. 4, D-80799 M\"{u}nchen, Germany}
\affiliation{Center for NanoScience ({\it CeNS}), Ludwig-Maximilians-Universit{\"a}t M{\"u}̈nchen, Geschwister-Scholl-Platz 1, D-80539 M\"{u}nchen, Germany}

\begin{abstract}

We explore the possibility to entangle an excitonic two-level system in a
semiconductor quantum dot (QD) with a cavity defined on a photonic crystal by
sweeping the cavity frequency across its resonance with the exciton transition.
The dynamic cavity detuning is established by a radio frequency surface acoustic
wave (SAW).  It induces Landau-Zener (LZ) transitions between the excitonic and
the photonic degrees of freedom and thereby creates a superposition state. We
optimize this scheme by using tailored Fourier-synthesized SAW pulses with up to
five harmonics. The theoretical study is performed with a master equation
approach for present state-of-the-art setups. Assuming experimentally demonstrated
system parameters, we demonstrate that the composed pulses increase both the
maximum entanglement and its persistence. The latter is only limited by the
dominant dephasing mechanism; i.e., the photon loss from the cavity. 

\end{abstract}


\date{\today}

\maketitle

\section{Introduction}

Entanglement, an intrinsically quantum mechanical correlation in composite
systems, is indispensable for most quantum information protocols
\cite{Nielsen2000a, Horodecki2009a} and, thus, should be available
for any qubit realization.  By definition, entanglement cannot be created by
local operations and, thus, requires some
interaction between the subsystems.  In order to achieve a controlled degree of
entanglement, one may turn on and off the effective interaction by tuning the subsystems
into or close to resonance for a limited time.  This includes a linear sweep
across the resonance giving rise to a Landau-Zener (LZ) scenario at an avoided
crossing.  In between the regimes of adiabatic following and sudden switching,
this process splits the wavefunction into two parts with a well-defined phase
and thereby creates an entangled state.

Since entanglement relies on a well-defined phase relation, it is fundamentally
limited by the susceptibility to decoherence of the chosen architecture. In the
very active field of solid state quantum systems the focus of LZ-based
entanglement creation was set in the past mainly to superconducting
\cite{Saito2006a, Wubs2007a} or spin-based \cite{Ribeiro2009a} setups using their
remarkable coherence properties. Although all-electrical radio frequency control
can be readily implemented in these systems, transfer of the encoded quantum
information to ``flying'' photonic qubits \cite{Cirac1997a, Wang2005a}
at optical frequencies is extremely
challenging. On the other hand, optically active QD nanostructures provide a
versatile platform allowing for tunable inter-dot coupling of exciton and spin
states and for coupling excitonic two-level systems to optical resonators to
implement a solid state cavity QED system. For such optically active systems the
implementation of LZ-schemes has been considered extremely challenging, because
most tuning mechanisms are quasi-static. While resonators available for cavity
QED allow for reaching the required regime of strong light-matter interaction,
the lifetime of cavity photons is typically more than two orders of magnitude
shorter than the coherence times of excitons in an isotropic environment. Thus,
theoretical proposals and experiments almost exclusively focused on quantum
operations using the non-linear optical properties of a static system.

\begin{figure}[b]
\includegraphics[width=\columnwidth]{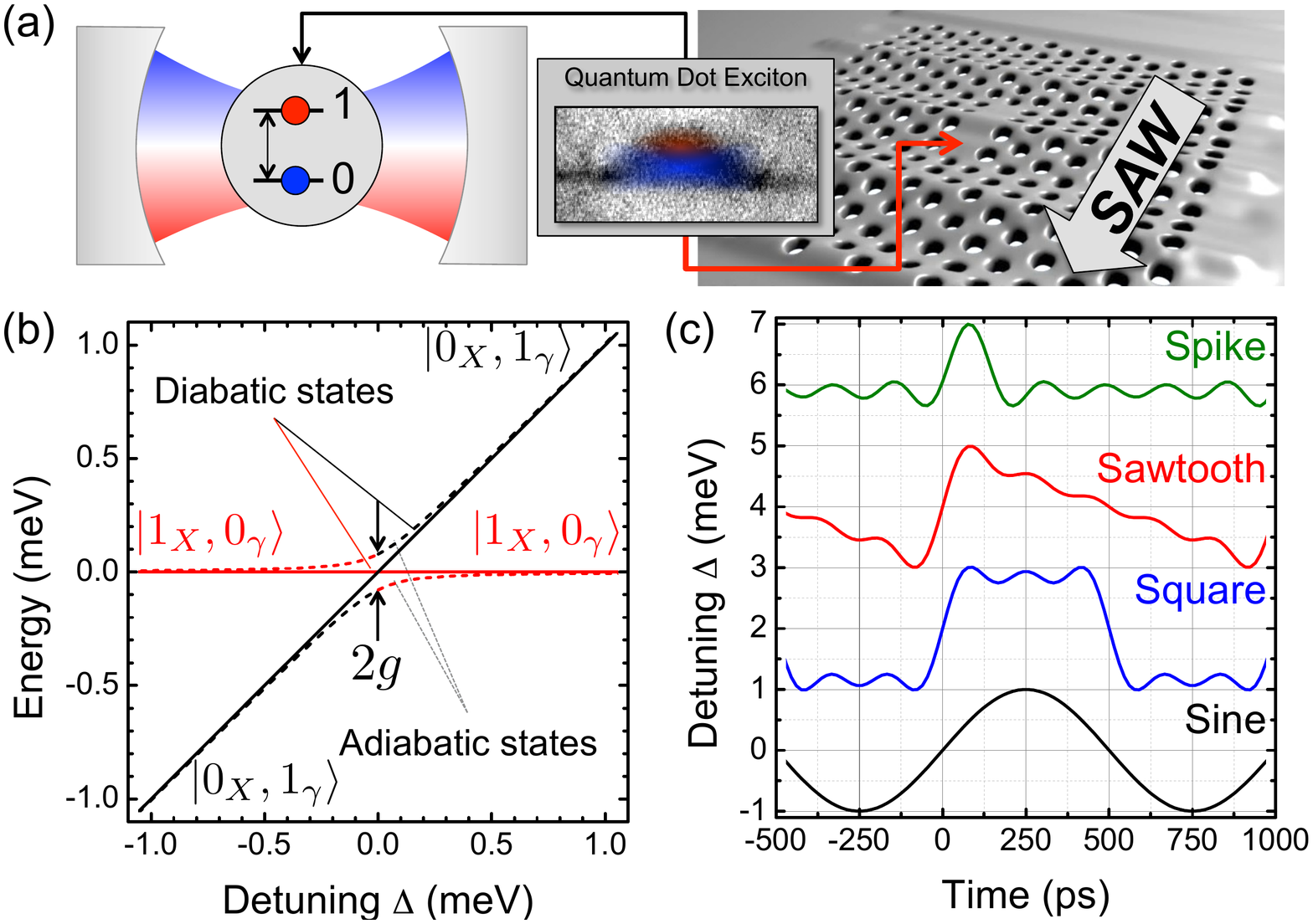}
\caption{(Color online) (a) Semiconductor quantum dot placed in a photonic crystal serves as
realization for cavity QED. Applying surface acoustic waves (SAW) to the
photonic crystal leads to a modulation of the cavity resonance frequency. (b) Energy
spectrum of the system in dependence on the detuning $\Delta$.  Owing to the
cavity-QD coupling $g$, the exciton energy (black solid line) and the one-photon
energy (red solid line) split at the degeneracy point $\Delta=0$ to form an
avoided crossing (dotted lines).  (c) Wave forms considered herein, which
originate from a superposition of higher SAW harmonics.}
\label{fig:1}
\end{figure}
In this paper we develop strategies to implement LZ-gates in a
semiconductor cavity QED system consisting of an excitonic two-level system in a
single semiconductor QD coupled to a cavity localized in the optical mode of a
photonic resonator, see Fig.~\ref{fig:1}.  The time-dependent detuning is
provided by SAWs for which we consider various feasible pulse shapes.  In order
to simulate the dynamics of the system, we numerically solve a master equation
including the dissipative effects of (spontaneous) QD decay and photon loss.
Moreover, we show that for experimentally demonstrated system parameters,
entanglement with a persistence of the photon lifetime can be achieved. Finally,
we discuss directions to reduce the experimental complexity due to the high
drive frequencies by using tailored Fourier-synthesized gating pulses.

\section{Quantum dot in a nanocavity}

For the nanocavity we assume a high-\textit{Q} nanophotonic defect resonator
defined in a two-dimensional photonic crystal membrane \cite{Akahane:03}
interacting with a two-level system in a single semiconductor QD.  This
two-level system is formed by the crystal ground state of the QD and its
fundamental optical excitation with one electron-hole pair, a single exciton $X$
\cite{Zrenner:02}.
Such systems have been studied over the past decade by several groups
\cite{Noda:07} who successfully demonstrated in key experiments both, the weak
\cite{Englund:05,Kress:05,Chang:06} and the strong
\cite{Yoshie:04,Hennessy:07,Englund:07,Laucht:09,Thon:09,Ota:11} coupling regime of
cavity QED. Most importantly, this type of nanocavity uniquely allows for a
dynamic and reversible spectral control of the optical mode at Gigahertz
frequencies by the coherent acoustic phonon field of a radio frequency SAW
\cite{Fuhrmann:11} which is crucial for the implementation of LZ-gates. We
emphasize that the QD transition is sensitive to the strain and electric field
of the SAW. Our previous experimental data \cite{Schulein2013} as well as other
studies \cite{Gell:08} on self-assembled $\rm InGaAs$ QDs suggest that the
bandwidth of this modulation is roughly three times smaller than that of the
cavity, $\gamma$. Therefore, we set the QD transition to be constant during the
acoustic cycle and treat the nanocavity resonance as time-dependent. A schematic
of this system is shown in Fig.~\ref{fig:1}(a).

For our modeling we restrict ourselves exclusively to \emph{experimentally
demonstrated system parameters} for an ${\rm InGaAs}$-based system with optical
transitions at $E_0=\hbar \omega_0= 1.3\,{\rm eV}$ ($\lambda_0=954\,{\rm nm}$,
$\omega_0/2\pi=314\,{\rm THz}$).  First, for SAW excitation we assume
$f_\text{SAW}\leq 5\,{\rm GHz}$. These frequencies are well below the highest
frequencies reported for bulk GaAs and GaAs-based suspended membranes of
$f_\text{SAW}\geq24\,{\rm GHz}$ \cite{Kukushkin:04} and $f_\text{SAW}>9\,{\rm
GHz}$ \cite{Takagaki2002}, respectively.  Second, for the coupled QD-cavity
system, the limiting constituent is the photon loss rate $\Gamma_{\gamma}$ from
the photonic crystal nanocavity. For our investigations we use $\hbar
\Gamma_{\gamma} = 25\,\mu{\rm eV}$ ($\Gamma_{\gamma} = 38\,\text{GHz}$),
compatible with values reported by Ota \textit{et al.} \cite{Ota:11}.  This
value corresponds to a cavity quality factor of $Q = \nu /
\Delta\nu=5.2\cdot 10^4$ and a photon life time of
$\tau_{\gamma}=Q/\omega_0=26\,{\rm ps}$. The decay rate of the QD exciton $\hbar
\Gamma_X = 0.2  \mu {\rm eV}$ ($\Gamma_{X} = 300\,\text{MHz})$
\cite{Laucht:09} is low when compared to $\Gamma_{\gamma}$.
Furthermore, we assume the system in the strong coupling regime with a vacuum
Rabi splitting $\hbar g > \hbar\Gamma_{\gamma}=25\,\mu{\rm eV}$. The energies of
the adiabatic and non-adiabatic states of this coupled system are plotted in
Fig.~\ref{fig:1}(b) of as a function of detuning
$\Delta$. The adiabatic energies show the characteristic avoided crossing between
\textit{X} and $\gamma$. We can write the corresponding states
$\ket{n_X,n_\gamma}$, where $n_X=0,1$ correspond to the two states of the
two-level system, while $n_\gamma$ refers to the cavity photon number.
For adiabatically slow sweeping, the exciton state $\ket{1_X,0_\gamma}$
transforms into a photon, $\ket{0_X,1_\gamma}$, by adiabatic following.
Increasing the sweep velocity, the initial states remains partially occupied
with a probability given by the LZ-formula \cite{Landau1932a, Zener1932a,
Stueckelberg1932a, Majorana1932a}
\begin{equation}
\label{PLZ}
	P_\text{LZ} = 1- \exp(-2\pi g^{2}/\hbar v),
\end{equation}
where the coupling $g$ determines the splitting at the degeneracy point
$\Delta=0$.  In the derivation of Eq.~\eqref{PLZ}, it was assumed that the
detuning $\Delta(t)=vt$ varies at a constant rate $v$ which for our SAW-based
approach is proportional to the SAW frequency and SAW amplitude, $v\propto f_{\rm
SAW}A$. Since we restrict ourselves to experimentally accessible parameters, we
keep the amplitude of the SAW modulation constant throughout the paper to
$A=\max(\Delta)=1\,{\rm meV}$, a base frequency of $f_0=1\,{\rm GHz}$, and
waveforms Fourier synthesized by superpositions of harmonics: In addition
to pure \textit{sine} drive at $f_{\rm SAW}=n f_0$, $n=1,3,5$, we investigate
the three synthesized waveforms summarized in Fig.~\ref{fig:1}(c). In
the limit $n\rightarrow \infty$, they converge to a
$\delta$-peak, henceforth referred to as \textit{spike}, and the two standard
waveforms of a \textit{square} and a \textit{sawtooth}, respectively.  For a realistic
modeling, these waves consist of the fundamental frequency $f_0$
and its first four harmonics. For the LZ-gate,
the system is initialized by a Rabi-oscillation from its ground state to the
excited state \textit{X} by a laser pulse \cite{Zrenner:02,Krenner:05a}.  To
ensure selective excitation of $X$ and to avoid initial photon population in the
cavity the spectral bandwidth of this excitation pulse has to be $\delta
E_\text{laser}\leq 250\,\mu{\rm eV}$ corresponding to a pulse length $\delta
t_\text{laser}\geq 20.7\,{\rm ps}$.  This in turn requires that the
initialization has to occur at $t_0>\delta t_\text{laser}$ before the system is
tuned through resonance at which the initial product state $\ket{1_X,0_\gamma}$
is converted to an entangled state.

\subsection{Excitonic quantum dot coupled to a SAW-driven cavity}

Without the SAW-driving, the system sketched in Fig.~\ref{fig:1}(a) is described
by the Jaynes-Cummings (JC) Hamiltonian \cite{Gardiner1991a}
\begin{equation}
   \label{eq:HamJC}
	H_\text{JC} = \frac{\varepsilon}{2}\sigma_{z} + \hbar\omega_0 a^{\dagger}a
	+ g (a\sigma_{+} + a^{\dagger}\sigma_{-}) ,
\end{equation}
where the pseudo-spin operators $\sigma_{z}$ and $\sigma_{\pm}$ describe the
excitonic quantum dot within a two-level approximation in the basis of the
ground state $|0_X\rangle$ and the one-exciton state $|1_X\rangle$ with the
energy splitting $\varepsilon$.  The bosonic operators $a$ and $a^\dagger$ refer
to the cavity with resonance frequency $\omega_0$, which is dipole coupled to the
quantum dot according to $g\sigma_x(a^\dagger+a)$.  Close to resonance, i.e.,
for $\varepsilon\approx\omega_0$, we can neglect the counter-rotating terms
$a\sigma_-$ and $a^\dagger\sigma_+$ to obtain the last term of the Hamiltonian
\eqref{eq:HamJC}.  Then the Hilbert space of the composed system discerns into
doublets spanned by the states $|1_X,n_\gamma\rangle$ and
$|0_X,(n+1)_\gamma\rangle$.  As function of the detuning $\Delta =
\varepsilon-\hbar\omega_0$, the eigenenergies of $H_\text{JC}$ form avoided crossings
of width $2\sqrt{n+1}g$ \cite{WallsMilburn1995a}, see Fig.~\ref{fig:1}(b).  The
central idea is to exploit the LZ dynamics at the avoided crossing of the lowest
doublet to entangle the quantum dot with the cavity, i.e., to reach a final
state $\propto |1_X,0_\gamma\rangle + e^{i\varphi}|0_X,1_\gamma\rangle$ with a
well-defined but possibly unknown and time-dependent phase $\varphi$.

The SAW modulates the cavity frequency so that it becomes time-dependent, i.e.
$\omega_0 \to \omega_0(t)$.  This implies that also the detuning gets modulated
with time: $\Delta \to \Delta(t)$.  For the case of a sinusoidal wave it is
$\Delta(t) = \Delta_0+A\sin[\Omega(t-t_0)]$, where the amplitude $A$ has to
exceed the static detuning $\Delta_0$ to pass through the avoided crossing with
a velocity of the order $v \sim A\Omega$.  This sweep velocity is of crucial
importance because, according to the LZ formula \eqref{PLZ},
it determines the probability for adiabatically following the ground
state.  Significant exciton-cavity entanglement requires $P_\text{LZ} \approx
1/2$.  On the other hand, it is desirable to slow down the modulation after the
entanglement is created, so that further state manipulations or 
a readout of the quantum state can be performed.
Therefore we like to exploit recent experimental achievements of a controlled
superposition of higher harmonics to the SAW and consider more generic waves
that lead to the detuning $\Delta(t) =\Delta_{0} +\sum_{n=1}^{N}A_n\sin[n\Omega
(t-t_{0})+\phi_n]$.  In an experiment both the amplitudes $A_n$ and the phases
$\phi_n$ can be controlled rather well, which enables a flexible design of the
pulses.  Here we consider, besides purely sinusoidal driving, also waves with
the characteristic shapes of a square, a sawtooth, and a spike.  We restrict
ourselves to the experimentally feasible case in which those waves are
approximated by a fundamental angular frequency of $\Omega = 2\pi\times 1$\,GHz
and its harmonics up to order $N=5$, i.e., we consider the drivings
\begin{equation}
\Delta(t) = \Delta_{0} +
  \begin{cases}
 A \sin[\Omega (t-t_{0})] & \quad \text{sine},\\[0.9em]
   \sum\limits_{n=0}^{5} \frac{A}{N}\cos[n\Omega (t-t_{0})] & \quad \text{spike},\\[0.9em]
   \sum\limits_{n=0}^{2} \frac{A}{2n+1}\sin[n\Omega (t-t_{0})] & \quad \text{square},\\[0.9em]
   \sum\limits_{n=1}^{5} \frac{A}{n}\sin[n\Omega (t-t_{0})] & \quad \text{sawtooth} ,
  \end{cases}
\label{pulses}
\end{equation}
sketched in Fig.~\ref{fig:1}(c).  Moreover, in order to
highlight the benefit of non-sinusoidal pulses, we also consider
pure sine waves with the angular frequencies $3\Omega$ and $5\Omega$.

\subsection{Decoherence and master equation}

Entanglement is a genuine quantum feature and, thus, is rather sensitive to
decoherence caused by the interaction with environmental degrees of freedom.  In
our case the latter are mainly the photonic modes $\nu$ outside the cavity.
Their influence can be modeled by the system-bath Hamiltonian $H_\text{env} =
\sum_\nu \hbar\omega_\nu a_\nu^\dagger a_\nu + Z \sum_\nu
\lambda_\nu(a_\nu^\dagger+a_\nu)$, where $\omega_\nu$ denotes the frequency of
mode $\nu$, while $\lambda_\nu$ is the coupling strength to a system operator
$Z$ which here is the cavity dipole operator $Z_\gamma = a^\dagger+a$.  For a
later continuum limit, we assume for the coupling the Ohmic spectral density
$J(\omega) = \pi\sum_\nu |\lambda_\nu|^2 \delta(\omega-\omega_\nu) \equiv
\pi\alpha_\gamma\omega/2$ with the dimensionless dissipation parameter
$\alpha_\gamma$ \cite{Leggett1987a, Hanggi1990a, Weiss1999a}.  Under the
condition that the environment is initially in a Gibbs state, we derive for the
reduced system density operator $\rho$ a Bloch-Redfield master equation
\cite{Redfield1957a, Blum1996a}.  Moreover, we take also dissipative transitions
of the quantum dot into account, which we model in the same way but with the
excitonic dipole moment $Z_X = \sigma_++\sigma_-$.  For consistency with the
system Hamiltonian \eqref{eq:HamJC}, we apply a rotating-wave approximation to
the master equation such that we finally arrive at \cite{Gardiner1991a,
WallsMilburn1995a}
\begin{align}
\label{eq:master}
 \frac{d}{dt}\rho = -\frac{i}{\hbar}[H_\text{JC}(t),\rho]
+ \mathcal{L}_\gamma(\rho) + \mathcal{L}_X(\rho),
\end{align}
where the Lindblad forms
\begin{align}
\label{eq:Lind}
\mathcal{L}_\gamma(\rho) ={} &\frac{\Gamma_\gamma}{2} (2 a \rho a^{\dagger} -
a^{\dagger} a \rho - \rho a^{\dagger} a ) , \\
\mathcal{L}_X(\rho) ={} &\frac{\Gamma_X}{2} (2 \sigma_- \rho \sigma_+ -
\sigma_+\sigma_- \rho - \rho \sigma_+\sigma_-) ,
\end{align}
describe cavity and quantum dot dissipation, respectively.  The decay rates
$\Gamma_\gamma$ and $\Gamma_X$ are determined by $J(\omega)$ evaluated at the
transitions frequencies of the dot-cavity Hamiltonian \eqref{eq:HamJC}.  Owing
to the smallness of the detuning
$|\Delta|\ll \varepsilon,\omega_0(t)$ and the smooth Ohmic spectral density, we
can ignore the energy shifts induced by the dot-cavity coupling $g$, so that we
can evaluate the spectral densities at the bare cavity and exciton frequency to
obtain $\Gamma_\gamma = \pi\alpha_\gamma\omega_0$ and $\Gamma_X =
\pi\alpha_X\varepsilon$, respectively.  Notice that we have assumed that the
environmental temperature is rather low, such that thermal excitation is
negligible.  Then the final state of the time-independent problem is the ground
state $|0_X,0_\gamma\rangle$, also beyond the rotating-wave approximation
\cite{Thingna2012a}.

\section{Entanglement dynamics}
\begin{figure}[t]
\centering
\includegraphics[width=0.5\textwidth]{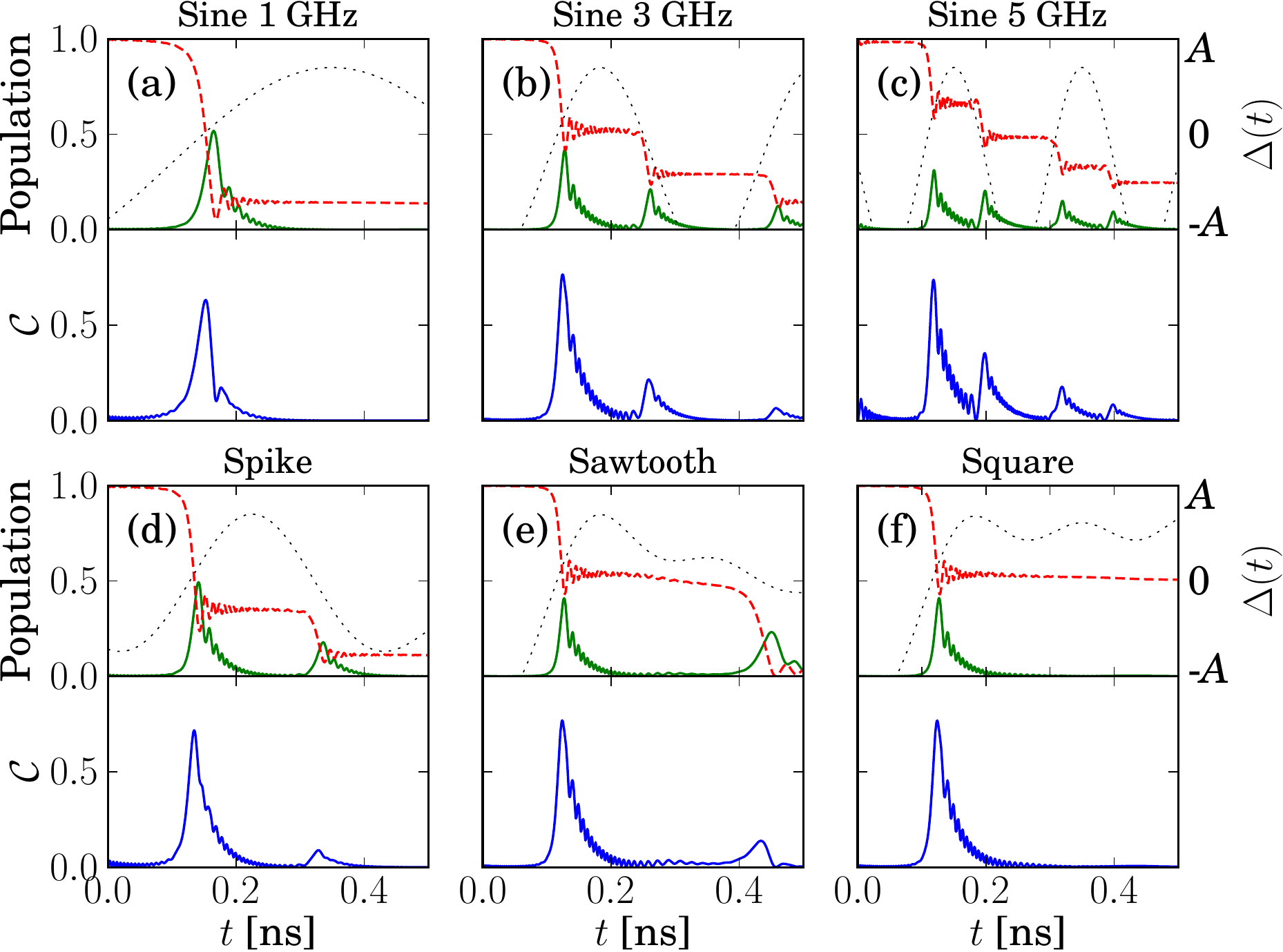}
\caption{(Color online) LZ entanglement dynamics for SAW shapes of a pure sine with frequencies
1\,GHz (a), 3\,GHz (b), and 5\,GHz (c), as well as for a spike (d), a sawtooth
(e), and a square (f), each with fundamental frequency $\Omega/2\pi = 1$\,GHz.
The cavity-dot coupling is $g=35\,\mu$eV, while the static detuning $\Delta_0 =
0.3$\,meV is modulated with an amplitude $A=1$\,meV, such that the crossing is
reached at time $t_0=T/10 = 0.5$\,ns.  The cavity and exciton decay rates read
$\Gamma_\gamma=25\,\mu\text{eV}/\hbar$ and $\Gamma_X = 0.2\,\mu\text{eV}/\hbar$.
Upper panels: Population of the states $|1_X,0_\gamma\rangle$ (red solid line)
and $|0_X,1_\gamma\rangle$ (green dashed line).  The dotted line visualizes the
course of the detuning $\Delta(t)$.
Lower panels: Cavity-dot entanglement in terms of the concurrence $\mathcal{C}$.
}
\label{fig:2}
\end{figure}
We consider the dynamics of the cavity-QD setup after an exciton is created at
time $t = 0$, while the cavity is empty, i.e., we numerically integrate the
master equation starting with the initial state $|1_X,0_\gamma\rangle$. In the
course of time, the SAW sweeps the energies of the two subsystems over an
avoided crossing with the state $|0_X,1_\gamma\rangle$ reached at $t\approx t_0$.
Moreover, the Lindblad terms cause a decay towards the ground state
$|0_X,0_\gamma\rangle$.  Our main aim is to investigate and to optimize the
degree of entanglement for differently shaped SAW pulses.  In order to quantify
the entanglement, we treat the cavity within two-level
approximation in the subspace spanned by the states $|0_\gamma\rangle$ and
$|1_\gamma\rangle$ with the corresponding Pauli matrices $\sigma_y^\gamma$.
This approximation is well justified, because our Hamiltonian \eqref{eq:HamJC}
preserves the total number of excitations, while our low-temperature dissipation
kernels $\mathcal{L}_\gamma$ and $\mathcal{L}_X$ only contain decay terms.  Then
we can employ as entanglement measure the concurrence \cite{Hill1997,Wooters1998a}
\begin{equation}
\label{eq:concurrence}
C(\rho) = \text{max}(\chi_{1}-\chi_{2}-\chi_{3}-\chi_{4},0),
\end{equation}
where the $\chi_{i}$ are the square roots of the eigenvalues of the matrix $\rho
\sigma_{y}\sigma_{y}^\gamma\rho^{*} \sigma_{y}\sigma_{y}^\gamma $ in
descending order. Here $\rho^{*}$ denotes the complex conjugate of the 
density matrix expressed in a basis of bell states \cite{Hill1997}.

In order to get a first impression of the dynamics, we depict in
Fig.~\ref{fig:2} the time evolution of the populations of the states
$|1_X,0_\gamma\rangle$ and $|0_X,1_\gamma\rangle$ and the corresponding
entanglement for various wave forms, while all other parameters are set equal.  For
a purely sinusoidal driving with frequency $\Omega/2\pi = 1$\,GHz, the
population of the initial state is by and large transferred to
$|0_X,1_\gamma\rangle$.  This corresponds to imperfect adiabatic following.  At
an intermediate stage at time $t\approx t_0$, the populations of both states are
comparable, while phase coherence between the participating states ensures good
entanglement with a concurrence up to $\mathcal{C}\approx 0.7$. However, since
for these parameters, $P_\text{LZ}$ is significantly larger than $1/2$, soon
after the crossing the one-photon state becomes highly populated.  Therefore the
systems disentangle soon after having passed the crossing.  Thus, we must
increase the sweep velocity, which can be achieved by using a higher frequency.
The results in panels (b) and (c), demonstrate that this can indeed augment the
concurrence.  Moreover, it increases the time during which the concurrence
exceeds a certain threshold value.  This ``entanglement persistence'' is mainly
limited by the cavity decay rate $\Gamma_\gamma$, at least under the realistic
condition $\Gamma_\gamma\gg\Gamma_X$.  Thus, our goal is to find parameters and
wave forms for which a significant entanglement is present during a time of the
order $1/\Gamma_\gamma$.  A theoretically interesting observation is that for
higher frequencies, the system may pass through the avoided crossing several
times in the time range considered.  The resulting repeated passages depend on
the phase acquired in between the crossings, leading to
Landau-Zener-St{\"u}ckelberg interference \cite{Zener1932a, Stueckelberg1932a,
Majorana1932a, Shevchenko2010a}.  However, for realistic cavity decay rates,
dephasing is too fast and, thus, the coherent superposition of our entangled
states turn into a (separable) mixture.  Therefore, we will not further discuss
interference effects.
\begin{figure*}[th!]
\centerline{\includegraphics[width=0.9\textwidth]{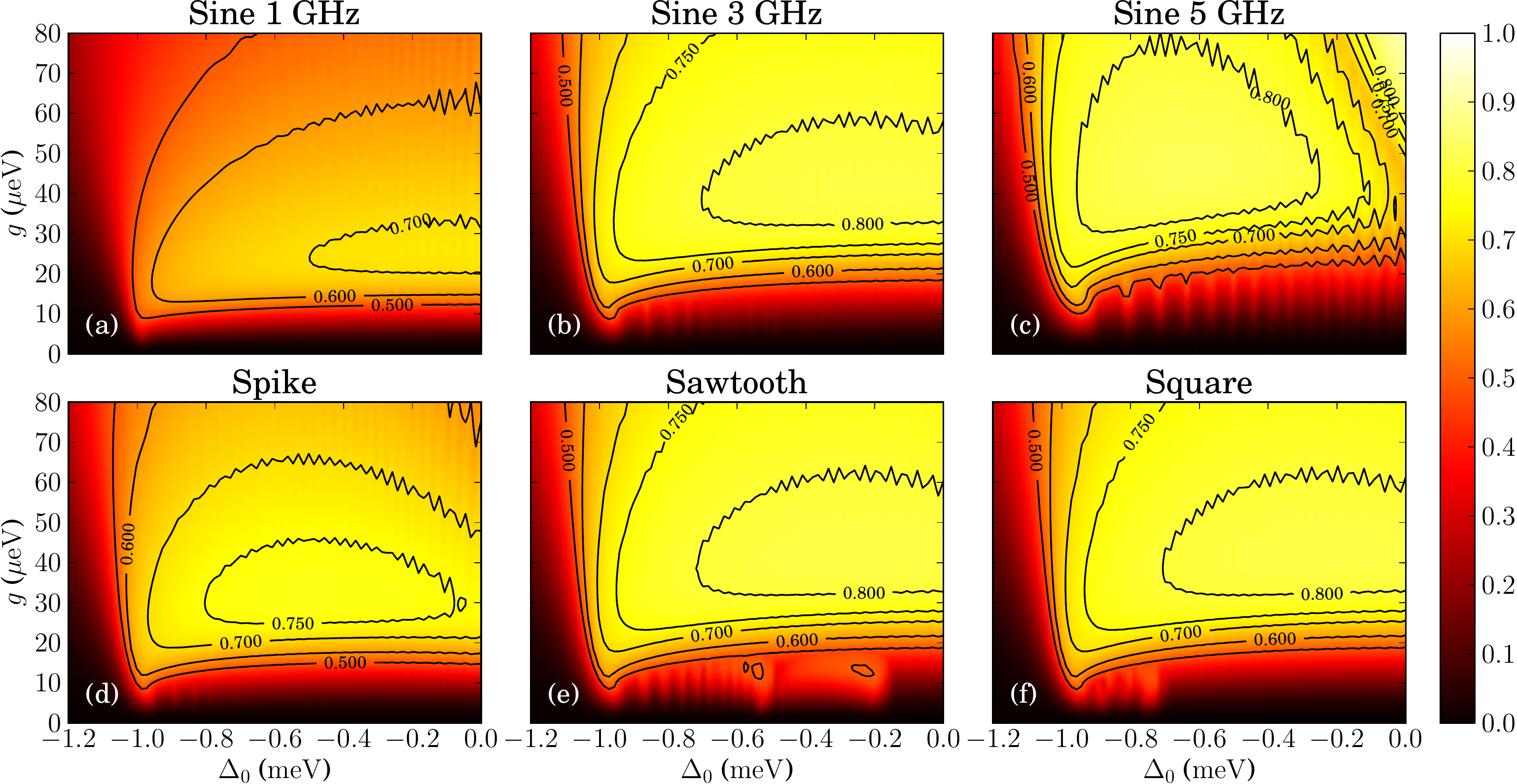}}
\caption{(Color online) Maximum of the concurrence achieved as function of the static detuning
$\Delta_0$ and the dot-cavity coupling $g$.  All other parameters, the wave
forms, and the arrangement of the panels is as in Fig.~\ref{fig:2}.}
\label{fig:3}
\end{figure*}
Even though SAWs with frequencies of 3 or 5\,GHz are feasible, inducing them
with a large intensity such that the detuning amplitude becomes 1\,meV
represents a rather difficult task. Waves with lower amplitudes are not
helpful, because they lead to a smaller sweep velocities and, thus, we would loose
what we gained from the higher frequency.  Moreover, the initial preparation of
the exciton must be performed during a fraction of the driving period,
because it takes a certain time and cannot be triggered with arbitrary
precision.  Thus for shorter driving periods, one will encounter difficulties to
carry out the preparation.  These difficulties can be circumvented by employing
more elaborate pulses such as the ones sketched in Fig.~\ref{fig:1}(c) and
mathematically expressed in Eq.~\eqref{pulses}.  Notice that for these pulses,
the contribution of each harmonic is significantly smaller than 1\,meV, while
the driving period remains at $2\pi/\Omega=1$\,ns.  The resulting entanglement
dynamics is plotted in Fig.~\ref{fig:2}(e--f).  As compared to panel (a), the
performance of the entanglement creation has improved.  Moreover, as we will see
below, this performance can be reached in a broader parameter range.

In an experimental implementation of our proposed scheme, one would on the one
hand like to obtain a rather large maximum for the concurrence, while on the
other hand, an appreciable entanglement should be found during a not too short
time, ideally limited only by the cavity decay $\Gamma_\gamma$.  Moreover, the
cavity-dot coupling $g$ is essentially a fixed parameter
determined during chip fabrication, which implies that the width of the avoided
crossing can be tuned only within a narrow range via the driving frequency and
the amplitude.  Therefore, it is desirable that the results do not depend too
sensitively on $g$.  Given this low flexibility, suggestions for more promising
SAW shapes are particularly welcome.

In order to characterize the performance of each wave form, we employ two
figures of merit.  The first one is the maximal concurrence $\mathcal{C}$
reached during a time $T/4$ centered at the avoided crossing as function of $g$
and $\Delta_0$, depicted in Fig.~\ref{fig:3}.  For all frequencies and pulse shapes
considered, the concurrence can reach values up to $\mathcal{C}\approx 0.8$.
However, the basic sinusoidal wave at 1\,GHz yields this value only in a
small range of the coupling $g$, which requires a precise fabrication process.
Since, as discussed above, for 3 and 5\,GHz, the required amplitudes of the
order 1\,meV are difficult to achieve, the more elaborate pulse shapes are
clearly preferable.  For all three composed pulses, the plots of the concurrence
maximum behave very similar.  The common feature of all three waves is the rather steep
slope of the detuning $\Delta(t)$ close to the center of the avoided crossing,
as can be appreciated in the upper panels of Fig.~\ref{fig:2}.  This suggests
that the main effect of the higher-order Fourier components is to augment the
sweep velocity at the crossing.

\begin{figure*}
\centerline{\includegraphics[width=0.9\textwidth]{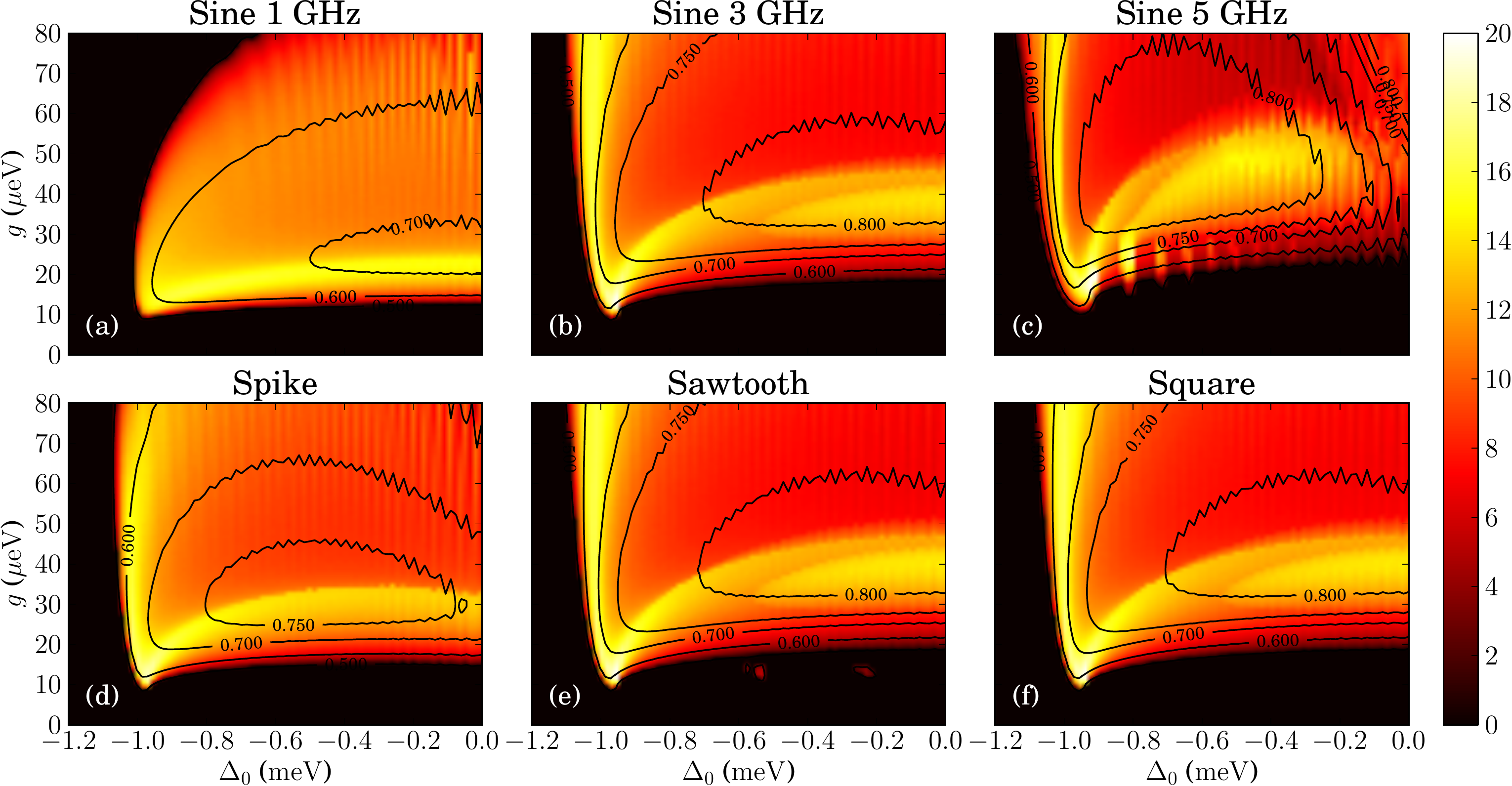}}
\caption{(Color online) Persistence of the entanglement, i.e., time during which the
concurrence exceeds the threshold value set by $\mathcal{C} > 1/2$.  All
parameters, wave forms, and the contour lines marking the maximum of the
concurrence are as in Fig.~\ref{fig:3}.}
\label{fig:4}
\end{figure*}
Our second figure of merit is the persistence time $\tau$ of the entanglement
defined as the time during which $\mathcal{C}>1/2$. This quantifier is depicted in
Fig.~\ref{fig:4}.  The contour lines enable a comparison with the results shown
in Fig.~\ref{fig:3}.  This reveals that a large maximum does not necessarily
coincide with long persistence.  This is particular the case for the sinusoidal
pulses with higher frequencies [panels (b) and (c)].  Nevertheless there exist
regimes where both the entanglement maximum and the persistence are rather
favorable and where $\tau$ practically reaches its theoretical limit, which is
the cavity lifetime $1/\Gamma_\gamma = 26$\,ps.  As for the maximum, the
plot for the sine wave with 3\,GHz and the ones for the composed pulses [panels
(d)--(f)] look similar.  However, the former has the disadvantage of being
experimentally more demanding.  Interestingly enough, in two regimes, albeit
small, pure sine waves yield surprisingly long entanglement duration:  First,
for 1\,GHz in the regime $\Delta_0\approx -1$\,meV, $g\gtrsim 10\,\mu$eV, where,
however, the maximum is rather low.  Second, for 5\,GHz, we witness in panel (f)
at $g\approx 25\,\mu$eV some islands with $\tau\gtrsim 20$\,ps. Combining the
two criteria of large maximum $\mathcal{C}$ and long persistence, we can
conclude that for an experimental realization, the quantum dot-cavity coupling
$g$ should be in the range 20--50\,$\mu$eV.

\section{discussion and conclusion}

We have preformed a theoretical study demonstrating the experimental feasibility
of entanglement generation in a semiconductor QD-nanocavity system by a
SAW-gated LZ-transition. Using exclusively experimentally demonstrated,
state-of-the-art system parameters we demonstrate high levels of entanglement
corresponding to a concurrence of $\mathcal{C}> 0.8$. Its persistence is mainly
limited by the photon loss from the cavity. This dominant dephasing mechanism
arises from the moderate quality factor of such semiconductor-based cavities. An
extension of this scheme to Fourier-synthesized SAW waveforms promises two
significant advantages over a single frequency sinusoidal drive. First, our
model predicts for square and sawtooth pulses a large concurrence over the system-limited
timescales for a broad range of $g$ and $\Delta_0$. The second advantage lies in
the experimental implementation: The complexity to achieve a sufficiently large
modulation amplitude of $1\,{\rm meV}$ increases significantly with increasing
$f_{\rm SAW}$, in particular for modulation frequencies of $3\,{\rm GHz}$ and
higher. In contrast,
Fourier-synthesized waveforms with only moderate amplitudes of higher harmonics
significantly overcompensate the additional requirements for the design
of the transducers for SAW generation. Moreover, the fundamental SAW period in
the experiment remains constant at $1\,{\rm ns}$ which facilitates the
synchronisation with the optical initialization and measurement of the
entanglement. The latter can be implemented, e.g., by extending existing schemes
based on reflectivity spectroscopy \cite{Englund:07} using short $(<1\,{\rm ps})$
and broadband laser pulses as a function of time during the acoustic cycle.
Moreover, since the system is initialized in the exciton state, i.e., in the lower
branch of the avoided crossing, any signal from the upper branch detected the
loss spectrum provides a fingerprint of the LZ-transition. Finally we want to
note, that the results of our theoretical study can be directly transferred to
other types of semiconductor cavities, most notably Bragg-type microcavities
\cite{Reithmaier:04}, which are SAW-compatible \cite{Lima2005}. In addition,
amongst the broad variety of control techniques, electrical tuning of the QD
transition via the quantum confined Stark
effect \cite{Laucht:09,Faraon:10,Kistner:09,Rakher:09} could be an
alternative approach to realize the required Gigahertz frequencies.

\begin{acknowledgments}
We gratefully acknowledge financial support by the Deutsche
Forschungsgemeinschaft (DFG) via Sonderforschungsbereich SFB 631 (projects A5
and B5) and the Emmy Noether Program (HJK, KR3790/2-1). This work was supported
by the Spanish Ministry of Economy and Competitiveness via through grant no.\
MAT2011-24331.
\end{acknowledgments}

\bibliographystyle{prsty}
\bibliography{new}

\end{document}